# Analytic solution of charge density of single wall carbon nanotube in conditions of field electron emission

## Zhibing Li and Weiliang Wang

The State Key Laboratory of Optoelectronic Materials and Technologies
Department of Physics, Zhongshan University, Guangzhou, 510275, China

**Abstract**

We derived the analytic solution of induced electrostatic potential along single wall carbon nanotubes. Under the hypothesis of constant density of states in the charge-neutral level, we are able to obtain the linear density of excess charge in an external field parallel to the tube axis.

The carbon nanotube (CNT) as a quasi one-dimensional material has been widely used as field electron emitter [1,2]. The set up of field electron emission is given in Fig.1 schematically. The absolute value of the applied field that parallel to the tube axis is denoted by $F_0$. When $F_0$ is applied between the cathode and anode, electrons will have opportunity to emit from the CNT apex to the vacuum by quantum tunneling. It is expected that the large aspect ratio would lead to a significant field enhancement at the apex surface and make the field emission much easier than usual metal plain emitters. There have been extensive simulations by ab initio methods [3,4,5,6]. However, the ab initio methods so far can only simulate short CNTs of few hundreds nanomenters. Some important features would have been missed in these simulations. Recently Refs. [7,8] have been succeeded in employing the linear-scaling divide-and-conquer (DAC) method [9] to simulate the charge distribution in CNTs of micrometer long. It is found that the excess charge distribution along the entire CNT has impact consequence on the field electron emission. The field enhancement is not the only story in CNT field emission. In [7,8], the excess charge density is determined by fitting the quantum simulation in a narrow window. It was time consuming and difficult to get stable results. The aim of the present paper is to derive the excess charge density analytically. With this solution, the simulation for field electron emission from CNTs of large size can greatly be accelerated.

We will consider a metallic single wall carbon nanotube (SWCNT) with constant density of states in the vicinity of its charge-neutrality level (CNL). The CNL is equal to the Fermi level for an isolated neutral CNT in the absence of applied field. It is also aligned with the Fermi level of the cathode if one ignores the Schottky barrier in the back contact. In the follows, we will farther assume that the system is in quasi-equilibrium so even in an applied field the Fermi levels of the CNT and cathode are aligned. Due to the applied field, some charges will move to the CNT from the cathode and accumulate in the CNT. The excess charge in turn induces an electrostatic potential to be denoted by $\varphi_I(z)$. The total electrostatic potential $\varphi_C(z)$ is the sum of $\varphi_I(z)$ and the applied potential ($zF_0$). Since the CNT is charged, CNL does not coincide with the Fermi level any more, except on the cathode (z=0). Relative to the Fermi level, the CNL is given by $-e\varphi_C(z)$. Denote the linear density of state in the vicinity of CNL as $N_f$ (it is assumed to be a constant). Then the linear excess charge is given by

$$\rho(z) = -eN_f[E_f + e\varphi(z)] = -e^2 N_f (F_0 z + \varphi_I(z)) \quad (1)$$

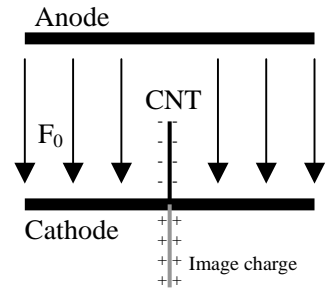

Fig. 1 Set up of field electron emission.

When the equilibrium is reached, from the Coulomb's law one has,

$$\varphi_I(z) = \int_0^L \int_0^{2\pi} \frac{\rho(z')}{4\pi\varepsilon} \frac{1}{\sqrt{(z-z')^2 + d^2 \sin^2(\theta/2)}} \frac{1}{2\pi} d\theta dz' \\ - \int_0^L \int_0^{2\pi} \frac{\rho(z')}{4\pi\varepsilon} \frac{1}{\sqrt{(z+z')^2 + d^2 \sin^2(\theta/2)}} \frac{1}{2\pi} d\theta dz' \quad (2)$$

Where d and L are the diameter and length of the SWCNT, respectively. The second term comes from the contribution of the image charge. The integration over the polar $\theta$ of the SWCNT can be carried out,

$$\varphi_I(z) = \int_0^L \frac{\rho(z')}{2\pi\varepsilon d} [G(z-z') - G(z+z')] dz' \quad (3)$$

The Green function is defined as

$$G(x) = \frac{K(-\frac{1}{x^2})}{|x|} \quad (4)$$

where K(s) is the complete elliptic integral of the first kind. Insert (1) into (3), we obtain a self-consistency relation for the induced electrostatic potential,

$$\varphi_I(z) = -M \int_0^L (z'F_0 + \varphi_I(z'))[G(z-z') - G(z+z')] dz' \quad (5)$$

where

$$M = \frac{e^2 N_f}{2\pi^2 \varepsilon d} \quad (6)$$

It is obvious that $\varphi_I(z)$ is an odd function of z. First let us consider a linear ansatz of the induced electrostatic potential,

$$\varphi_I^0(z) = a_1 z \quad (7)$$

where $a_1$ is a parameter to be fixed. Inserting (7) into the r.h.s. of (5), one can obtain

$$\varphi_{I0}(z) = -Md(F_0 + a_1)\Gamma_0(z) \quad (8)$$

where

$$\Gamma_0(z) = (L-z)E(-x_-^{-2}) - (L+z)E(-x_+^{-2}) \\ + \frac{\pi z}{16} \left[ \frac{4 F_3(-x_-^{-2})}{x_-^2} + \frac{4 F_3(-x_+^{-2})}{x_+^2} + 4Log(x_- x_+) - 2C \right] \quad (9)$$

In (9), $E(s)$ is the complete elliptic integral of the second kind, $_4F_3(s)$ is the brevity of the generalized hypergeometric series,

$$_4F_3(s) = {}_4F_3(\{1,1,3/2,3/2\},\{2,2,2\},s)$$

We have defined

$$x_\pm = \left(\frac{L \pm z}{d}\right)^2 \quad (10)$$

and a numerical integration constant $C = -11.09035$. Expanding $\Gamma_0(z)$ at z=0, one has,

$$\Gamma_0(z) = b_1 z + b_2 z \ln\left(\frac{L^2 - z^2}{L^2}\right) + \cdots \quad (11)$$

where $b_1$ and $b_2$ are constants. One has

$$b_1 = -2K(-\frac{d^2}{L^2}) \\ + \frac{\pi}{8} \left[ \frac{d^2}{L^2} \cdot {}_4F_3(-\frac{d^2}{L^2}) + 4Log(\frac{L^2}{d^2}) - C \right\} \quad (12)$$

Equate (7) and (8) and keep only the linear term, we obtain

$$a_1 = -\frac{MdF_0 b_1}{1 + Mdb_1} \quad (13)$$

The expansions (11) suggest to improve the ansatz for the induced electrostatic potential as

$$\varphi_I^1(z) = a_1 z + a_2 z \ln\left(\frac{L^2 - z^2}{L^2}\right) \qquad (14)$$

where $a_2$ is a new parameter to be fixed by the self-consistency equation (5). The singularity at $z = L$ is because the failure of classical theory in the atomic scale. Eq. (14) is only valid for the column of the SWCNT. In the tip regime (within 100 nm from the end of SWCNT), the quantum mechanical calculation is necessary. Insert (14) to the r.h.s. of (5), one will obtain an improved approximate solution for the induced potential $\varphi_{I1}(z)$. Since the integration is lengthy, we will only give the numerical result. For a (5,5) SWCNT, d=0.7 nm. We adopt the linear density of state in CNL of the tight binding theory [10], $N_f$=2.2 (nm·eV)$^{-1}$. M turns out to be 2.00136. For L=1000 nm, we obtain

$$a_2 = 0.0011 F_0 \qquad (15)$$

In Fig.2, we plot the $\varphi_{I0}(z)$, $\varphi_{I1}(z)$, and $\varphi_I^1(z)$ (from uppermost, middle, to lowest curve, respectively) in the applied field $F_0 = 10 V/\mu m$. It can be seen that the improved ansatz satisfies the self-consistency relation quite good in the column of CNT.

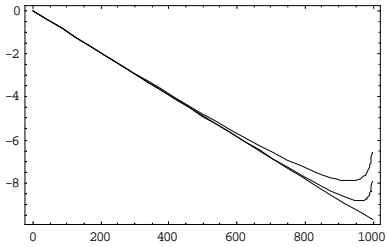

Fig. 2 $\varphi_{I0}(z)$, $\varphi_{I1}(z)$, and $\varphi_I^1(z)$ in $F_0$=10 V/μm, from uppermost, middle, to lowest curve, respectively. The horizontal axis is in nanometer, the vertical axis is in V/μm.

The linear excess charge density $\rho(z)$ given by (1) is shown in Fig. 3 for the same applied field of Fig.2. The straight line is obtained via the linear ansatz. The curve bending up is the result of the improved ansatz. One can see that the linear excess charge density can be well approximated by a linear function.

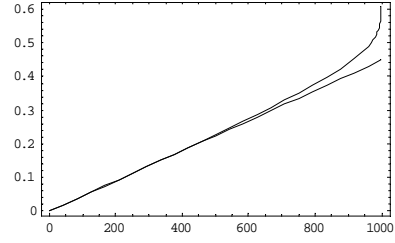

Fig.3 The linear excess charge density (in e/nm). The straight line is calculated via the linear ansatz, the curve bending up is via the improved ansatz.

Note that the way we find the ansatz of induced potential is a systematic procedure to generate a series that asymptotically approaches the exact solution. The idea to construct the ansatz is as following. Since the excess charge is induced by the uniform external field, the leading term of induced potential must be a linear potential. This is consistent with the result of Ref. [11]. In the hypothesis of constant density of states, the linear potential leads to a linear charge distribution. However, from the integral of (5) one can see that the linear charge distribution can induce nonlinear terms for the electrostatic potential. For the consistency of (5), the nonlinear terms generated by the integral should be included in the ansatz of the potential. Thereby we have an improved ansatz for the potential. The new ansatz will generate more terms in the integral of (5). So the ansatz can be farther improved by include those terms. In principle, the iteration can go on until a satisfactory solution is obtained. In the present paper, we keep the leading terms in the expansion of z/L and the singularity term generated by the second iteration. That should be enough for long CNT.

In conclusion, we obtain the electrostatic potential along SWCNTs and the excess charge density by a systematic method. Our result shows that the band bending (its amound is equal to the electrostatic potential) can not be ignored as most calculations of tight-binding method and density functional theory have done.

Authors thank the valuable discussions with J. Peng, G.H. Chen, C. Edgcombe, and R. Forbes. The research is supported by the projects of the National Natural Science Foundation of China (the Distinguished Creative Group Project; Grant No. 90103028, 90306016).